# Diamond nanothread as a new reinforcement for nanocomposites

Haifei Zhan[1], Gang Zhang[2,*], Vincent BC Tan[3], Yuan Cheng[2], John M. Bell[1], Yong-Wei Zhang[2], and Yuantong Gu[1,**]

[1]*School of Chemistry, Physics and Mechanical Engineering, Queensland University of Technology (QUT), Brisbane QLD 4001, Australia*
[2]*Institute of High Performance Computing, Agency for Science, Technology and Research, 1 Fusionopolis Way, Singapore 138632*
[3]*Department of Mechanical Engineering, National University of Singapore, 9 Engineering Drive 1, Singapore 117576*

*Corresponding authors:*
*Dr Gang Zhang, zhangg@ihpc.a-star.edu.sg
**Prof Yuantong Gu, yuantong.gu@qut.edu.au

**Abstracts:** This work explores the application of a new one-dimensional carbon nanomaterial, the diamond nanothread (DNT), as a reinforcement for nanocomposites. Owing to the existence of Stone-Wales transformation defects, the DNT intrinsically possesses irregular surfaces, which is expected to enhance the non-covalent interfacial load transfer. Through a series of *in silico* pull-out studies of the DNT in polyethylene (PE) matrix, we found that the load transfer between DNT and PE matrix is dominated by the non-covalent interactions, in particular the van der Waals interactions. Although the hydrogenated surface of the DNT reduces the strength of the van der Waals interactions at the interface, the irregular surface of the DNT can compensate for the weak bonds. These factors lead to an interfacial shear strength of the DNT/PE interface comparable with that of the carbon nanotube (CNT)/PE interface. Our results show that the DNT/PE interfacial shear strength remains high even as the number of Stone-Wales transformation defects decreases. It can be enhanced further by increasing the PE density or introduction of functional groups to the DNT, both of which greatly increase the non-covalent interactions.

# 1. Introduction

Low-dimensional carbon nanomaterials have been widely used as reinforcements for nanocomposites or bio-molecules to enhance their mechanical and thermal performance.[1-3] For mechanical applications, the interfacial shear strength is a crucial factor that determines the effectiveness of load transfer between the polymer matrix and nanoreinforcements.[4] In principle, the interfacial shear strength is determined by the interfacial bonds between the reinforcements and the polymer matrix. Various approaches have been proposed to enhance the interfacial bonding, including defect functionalization, non-covalent and covalent functionalization.[5-6] As an example, significant research has been devoted to create functionalized carbon nanotube (CNT) surface or CNT-based nanostructures in order to establish strong covalent bonds with the polymer chains.[7] Recent work[8] has reported the decoration of CNT with polymer nanocrystals, which results in a so-called nanohybrid shish kebab (NHSK) structure. The NHSK structure has a rougher surface, providing a non-covalent approach for the enhancement of the interfacial interactions between CNT and polymer matrix.[4]

Motivated by this, the very recently reported one-dimensional diamond nanothread (DNT), as synthesized through solid-state reaction of benzene under high-pressure,[9] has potential as a good reinforcement for nanocomposites. Previous work has shown that the DNT has excellent mechanical properties, and in particular a high stiffness of about 850 GPa.[10] DNT is a $sp^3$-bonded carbon structure, and it can be regarded as hydrogenated (3,0) CNTs connected with Stone-Wales transformation defects (see inset of Figure 1a).[11] However, unlike the perfect structure of CNTs, the existence of SW transformation defects interrupts the central hollow of the structure and introduces irregular surfaces (inset of Figure 1a). Also, the structure of the DNT can be further altered by changing the connectivity of the constituent carbon rings.[12-13] Together with the hydrogen exposure surface of the DNT, that is easy to chemically modify, such imperfections are expected to enhance the interfacial load transfer, and make the DNT an ideal reinforcement candidate for nanocomposites. Here, load transfer is defined as the load bearing ability of the interface, which is related to the energy required to "break" the interface and pull-out the DNT.

To achieve the technological potential of DNT-reinforced nanocomposites, there are several questions that need to be answered: What is the underlying mechanism that governs the load transfer between DNT and polymer? How can the interfacial load transfer be enhanced? To answer these questions, this work carried out a series of *in silico* studies, focusing on the pull-out behavior of the DNT inside a representative polyethylene (PE) polymer matrix. We found that the load transfer between DNT and PE matrix are dominated by van der Waals interactions, and the DNT-PE interface has an interfacial shear strength similar to that of the CNT in a PE matrix. We also demonstrated that by increasing the PE density, or by introducing functional groups on the DNT, the interfacial shear strength can be increased.

## 2. Results and Discussions

High-density polyethylene (PE) was chosen to build the representative polymer model. For simplicity, we adopt a pristine linear polyethylene molecule, and the model was constructed through the packing module[14] using Materials Studio (version 6.0) from Accelrys Inc. A sample DNT (length of ~ 5.3 nm) with four Stone-Wales transformation defects (SWDs) was first established based on the recent experimental results and first principle calculations.[9] For discussion convenience, the DNT with *n* SWDs is denoted as model DNT-*n*, e.g., the DNT-4 has four SWDs (inset of Figure 1a). The DNT model was then put into the middle of an initial periodic box with the width and length of 40 Å, and height of $h$ Å. Here, the height $h$ is about 6 nm, which is the length of the embedded DNT. Prior to packing, a van der Waals (vdW) surface was created on the DNT surface (inset of Figure 1a) to enclose the available filling space. Afterwards, the PE model is generated by filling the box with chains containing 20 monomers, i.e., $-(CH_2-CH_2)_{20}-$, with an experimentally measured density of 0.92 g/cm$^3$ (Figure 1a).[15] During the packing process, the polymer consistent force field (PCFF)[16] was used to describe atomic interactions within the polymer and the DNT, which has been shown to reproduce well the mechanical properties, compressibility, heat capacities and cohesive energies of polymers and organic materials. For the vdW interactions, the 6-9 Lennard-Jones (LJ) potential was applied with a cutoff of 12.5 Å, and the Ewald summation method was employed to treat the long-range Coulomb interactions.[17]

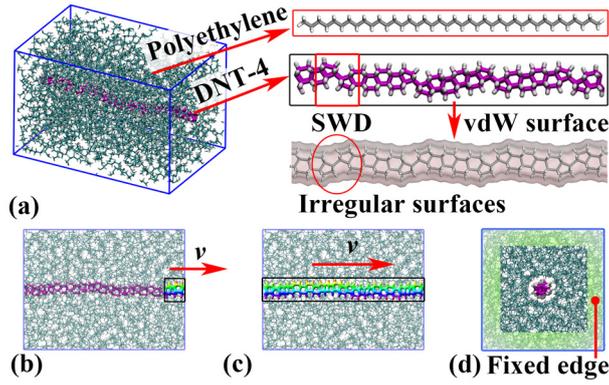

**Figure 1** (a) A periodic PE composite model containing DNT-4. The inset shows the atomic configurations of the polyethylene, DNT-4 and also its van der Waals surface; Schematic view of the pull-out setting with the velocity load applied to: (b) the end of the DNT, and (c) the whole rigid DNT, the loading region of the DNT is treated as a rigid body; (d) Cross-sectional view of the PE composite shows the fixed edges during pull-out simulation.

*2.1 Load transfer between DNT and PE*

Initially, we consider the pull-out of a DNT with four SWDs (i.e., DNT-4) from the PE matrix to exploit the load transfer mechanisms (periodic model size of 40 × 40 × 53 Å$^3$). A constant velocity is applied to one end of the DNT (which is treated as a rigid body) with the rest as a flexible body (Figure 1b), such scheme closely mimics the experimental setup.[18-19] As illustrated in Figure 2a, the total energy change ($\Delta E_{po}$) increases with the sliding distance and saturates at a value of ~ 9.5 eV when the DNT is fully pulled-out from the PE matrix. Note that Figure 2a shows a gradual decrease of $\Delta E_{po}$ after full pull-out due to the truncated sliding distance (i.e., simulation time), it actually fluctuates around 9.5 eV with continuing simulation time (up to 3.6 ns, sliding distance 180 Å. see Supporting Information).

Theoretically, the potential energy of the whole system comprises the sum of all the covalent bonds, van der Waals and Coulombic interactions. Specifically, under the PCFF force field, there is no bond breaking or disassociation either in the polymer matrix or the embedded DNT during the pull-out. In other words, the total energy change $\Delta E_{po}$ is mainly attributed to the change of the vdW and electrostatic energy. As shown in Figure 2a, the total vdW energy change ($\Delta E_{vdW}$) is similar to the total energy (~ 9.5 eV). This result suggests that for polymers without significant charges, the electrostatic interaction is very weak compared with the vdW interaction at the DNT/PE interface.

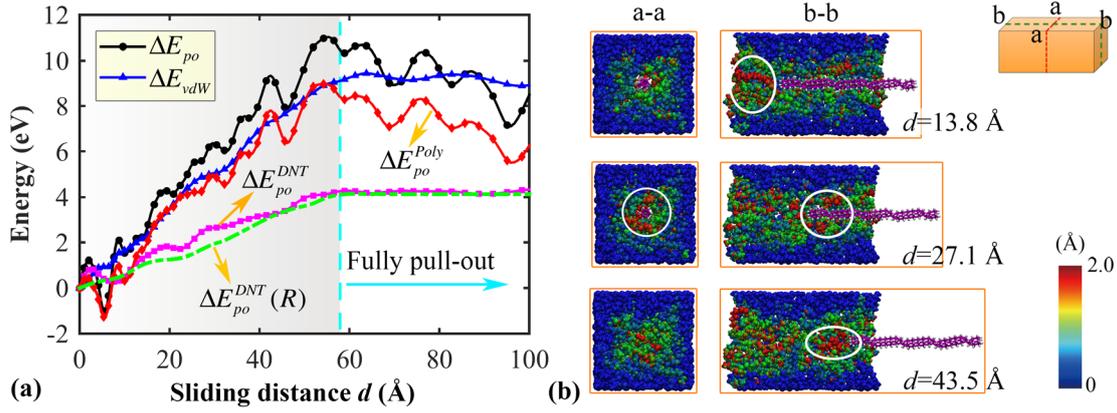

**Figure 2** (a) The potential energy change of the whole PE composite ($\Delta E_{po}$), the polymer matrix ($\Delta E_{po}^{Poly}$) and the DNT ($\Delta E_{po}^{DNT}$), and the total vdW energy change ($\Delta E_{vdW}$), as a function of the sliding distance. The potential energy change for the DNT while it is taken as a rigid body is also presented for comparisons ($\Delta E_{po}^{DNT}(R)$); (b) Cross-sectional views of the polymer atomic configurations at different sliding distance. The atoms are coloured according to the absolute relevant atomic displacements calculated according the adjacent atomic configurations. Atoms are coloured as red for displacements over 2 Å.

The total energy change, $\Delta E_{po}$, shows significant fluctuations, which result from the local deformation or relaxation of the polymer matrix. As evidenced in Figure 2a, these fluctuations are only seen in the profile of the total potential change of the polymer matrix, and the energy curve for the DNT ($\Delta E_{po}^{DNT}$) does not show significant fluctuations. During the pull-out, the attractive force between the embedded DNT and PE will induce local deformation to the adjacent PE chains. Also, along with the pull-out of the DNT, the release of the pre-occupied space also introduces a free surface inside the polymer, and thus allows a free relaxation of the polymer matrix that surrounded the DNT tail. Such relaxation can be readily seen from the absolute relevant atomic displacements in the composite structure. Here, the relevant atomic displacements are estimated every 2 Å sliding distance between the adjacent atomic configurations. As plotted in Figure 2b, the PE atoms that are surrounding the DNT, left behind after pull-out of the DNT, and located at the two free ends of the structure normally show a much larger displacement (highlighted by the circles). Thus, the absorption and release of the strain energy that accompanies these atomic displacements results in the local fluctuation of the total potential energy.

Of interest, we also investigate the pull-out behavior by treating the DNT as a rigid body (Figure 1c), which has been widely adopted for the pull-out simulation of CNT from polymer matrix.[20-22] A similar total potential energy change (~ 9.5 eV) is observed. In particular, the profile of the potential energy change of the DNT, $\Delta E_{po}^{DNT}(R)$, almost overlaps with that obtained from the case with non-rigid DNT (see Figure 2a), indicating that the deformation of the DNT is insignificant during the pull-out process. In all, the results suggest that the load transfer at the DNT/PE interface is dominated by the vdW interactions.

*2.2 The interfacial shear strength*

Since the DNT has a hydrogenated surface (inset of Figure 1a), the interfacial vdW interactions are weaker than those in CNT (with C surface). Thus, it is supposed that the DNT might have lower interfacial shear strength compared to CNT. The following section compares the interfacial shear strength at the PE/DNT and PE/CNT interfaces.

Following previous works,[22-23] the load transfer at the composite and reinforcement interface can be quantified by estimating the interfacial shear strength (ISS, $\tau$) based on either the total potential energy change ($\Delta E_{po}$) or the total vdW energy ($\Delta E_{vdW}$) during pull-out. Simplifying the PE matrix as a continuous medium and the DNT as a solid cylinder with a diameter of $D$, $\Delta E_{po}$ can be approximated as,

$$\Delta E_{po} = \int_0^L F(x)dx \qquad (1)$$

where $x$ is the sliding distance. The diameter is approximated as the distance between exterior surface hydrogens, i.e., ~ 0.5 nm, following Roman *et al.*[10] $F(x)$ is the shear force, which can be calculated as $F(x) = \pi D(L-x)\tau$. Here $L$ is the embedded length of the DNT. Thus, the $\tau$ can be written as

$$\tau = 2\Delta E_{po} / \pi D L^2 \qquad (2)$$

For the results presented in Figure 2a, $\Delta E_{po}$ is approximately $\Delta E_{vdW}$, i.e., 9.5 eV, and $L$ is about 5.8 nm, which yields to an ISS of ~ 58 MPa at the PE/DNT-4 interface.

In comparison, the pull-out of an ultra-thin (4,0) CNT from the PE matrix was simulated. As illustrated in Figure 3, the PE-CNT composite shows a total potential energy change around 12 eV, larger than that of the DNT-4. The total sliding distance is about 5.9 nm. Approximating the (4,0) CNT diameter as 0.65 nm (the summation of the inner radius of the CNT and the graphite interlayer distance 0.34 nm), an ISS about ~ 54 MPa is obtained. Such value is comparable with that reported for the interfacial shear strength at the PE/CNT(10,10) interface[24] (about 33 MPa) and PE/CNT(10,0) interface[20] (about 133 MPa). These initial calculations have suggested that despite the hydrogenated surface, the DNT has comparable interfacial shear strength compared to CNT (see Supporting Information for the discussion of ISS with different sliding distance approximations).

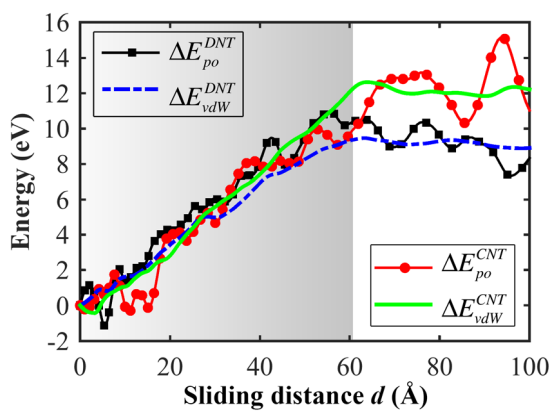

**Figure 3** Comparisons of the total potential energy change, total vdW energy change between the composites embedded with DNT-4 and (4,0) CNT.

Compared with the excellent tensile properties of DNT (tensile stiffness approaching 850 GPa and yield strength around 26.4 nN),[10] the estimated ISS (~ 58 MPa) is very small, which explains our observation that the deformation of the DNT is insignificant during the pull-out. However, the ISS is comparable to that of CNT, which is unexpected based on the nature of the surfaces. In the DNT-PE composite, however, the presence of the SW transformation defects is expected to lead to the polymer matrix experiencing lateral stresses which are generated as the polymer chains are pushed away from the DNT during the pull-out. This is in addition to the shear stresses along the interface. Taking the deformation of the PE matrix in normal and interfacial (the pull-out) directions, and summing them up into an effective interfacial shear deformation leads to enhanced shear strength. In other words, the weaker vdW interaction at the PE/DNT interface is in effect assisted by mechanical interlocking, and therefore leads to a comparable ISS with that of the PE/CNT interface. We note that the hydrogenated surface of DNT would also lead to different topologies of the PE compared with the CNT. However, how different topologies of the PE would affect the interfacial strength still requires further study.

*2.3 Influential factors on the interfacial shear strength*

The discussion above has shown that the DNT/PE interfacial shear strength is influenced by both the vdW interactions and local deformation of the polymer matrix during sliding. In this section, we assess how the interfacial shear strength can be engineered. Since the total potential energy change normally exhibits relatively large fluctuations and approximates to the total vdW energy change, following estimations of the ISS will use the total vdW energy change $\Delta E_{vdW}$.

Considering the tunability of the DNT's structure, we first compare the pull-out of the embedded DNT contains one SWD (DNT-1) and two SWDs (DNT-2). The corresponding periodic polymer model sizes are $40 \times 40 \times 61$ Å$^3$ and $40 \times 40 \times 60$ Å$^3$, respectively. As shown in the left panel of Figure 4a, a similar ISS is obtained for all three studied DNTs, indicating an insignificant influence from the number of SWDs. In comparison, as the interfacial shear strength is principally determined by the vdW interactions, one straightforward enhancement approach is to increase the interactions sites at the interface. In this respect, we consider the PE composite (embedded with DNT-2) with a higher initial density of 0.95 and 1.00 g/cc based on the experimental measurements.[15] As expected (middle panel of Figure 4a), the ISS increases when the PE density increases from 0.92 to 1.00 g/cc.

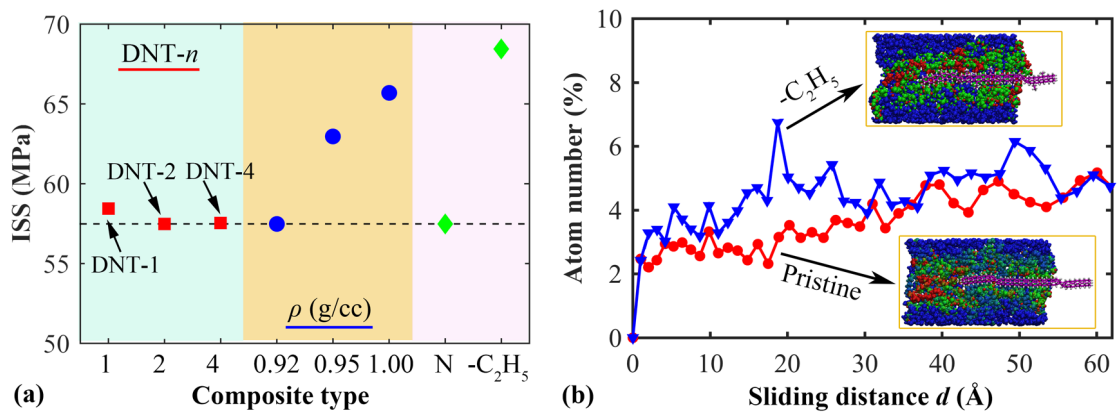

**Figure 4** (a) Estimated ISS for PE composites with different DNTs (DNT-1, DNT-2 and DNT-4, left panel), with different initial density (0.92, 0.95 and 1.00 g/cc, middle panel), with pristine DNT-2 (N) or –C$_2$H$_5$ functionalized DNT-2 (right panel). Refer to the Supporting Information for the trajectory of $\Delta E_{vdW}$ for these studied samples; (b) The percentage of PE atoms with atomic displacements over 2 Å. Insets show the atomic configurations at the sliding distance of ~ 19 Å, and atoms with atomic displacements over 2 Å are colored red.

Another avenue to enhance the load transfer is to introduce functional groups to the embedded DNTs. In the literature, a variety of functional groups on single-wall CNT or graphene have been discussed, such as –COOH, -CONH$_2$, -C$_6$H$_{11}$ and –C$_6$H$_5$.[24-26] To study this effect, we randomly replaced 5% of the DNT-2's surface H atoms by -C$_2$H$_5$ molecules. It is found that the total vdW energy change increased from ~ 10.5 eV to ~ 12.5 eV due to the presence of the -C$_2$H$_5$ functional groups, which corresponds to about 20% increase in ISS (from ~ 58 MPa to ~ 68 MPa). It is expected that introducing functional groups to DNT will not only increase the vdW interaction sites between DNT and PE matrix, but also result in more local deformation of polymer during pull-out. This is confirmed in Figure 4b which shows the trajectory of the fraction of PE atoms with atomic displacements over 2 Å. More PE atoms in the composite with functionalized DNT experience large atomic displacements during pull-out than for the composite with pristine DNT, suggesting more severe local deformation of PE matrix. In all, it is found that the number of SWDs exerts insignificant influence to the ISS, whereas, the polymer density as well as the functional groups can effectively enhance the load transfer at the interface.

## 3. Conclusions

Based on large-scale atomistic simulations, this study assessed the potential of the novel one-dimensional DNT as reinforcement for polymer. The load transfer and interfacial shear strength are analyzed by considering a representative PE composite through pull-out simulation. It is found that the load transfer between DNT and PE matrix are largely dominated by the vdW interactions. Compared with CNT, the hydrogenated surface of the DNT weakens the vdW interactions at the interface. The irregular surfaces associated with the existence of SWDs helps to compensate such weakness, and therefore, the DNT/PE interface shows comparable interfacial shear strength with that of the CNT/PE interface. Further examinations show that the interfacial shear strength has a weak relationship with the number of the SWD of the embedded DNT. However increasing the PE density or introducing functional groups, can effectively enhance the interfacial shear strength. This study provides a first understanding of the load transfer between DNT and polymer, which will provide a basis for future implementation of DNTs as reinforcements for various nanocomposites. We note that current work only considered the non-covalent interface between DNT and polymer. However it is relatively easy to introduce covalent interactions, or even cross links with other DNT threads, at the hydrogenated surface of DNT. This could create multi-thread structures, and it is assumed to greatly enhance the load transfer and requiring further investigation.

## 4. Methods

The load transfer between DNT and polymer is acquired through a series of pull-out simulations carried out by the software package LAMMPS,[27] mimicking the pull-out experiments of CNTs from polymer matrix.[18] The length direction of the initial model was switched to non-periodic boundary conditions (the two lateral directions were still under periodic boundary conditions). Followed previous settings in Material Studio, the bonded interactions (within the polymer and the DNT) were described by the polymer consistent force field (PCFF)[16] and the vdW interactions were described by the LJ potential (with a cutoff of 12.5 Å). The Ewald summation method was employed to treat the long-range Coulomb interactions.[17] The polymer models were firstly subjected to an energy minimization using conjugate gradient method. Afterwards, two steps of relaxations were performed to arrive an equilibrium structure using Nosé-Hoover thermostat,[28-29] including the relaxation under constant pressure (NPT) ensemble for 200 ps (pressure of one atmosphere) with the DNT being fixed as a rigid body, and then the relaxation under constant temperature (NVT) ensemble for another 200 ps with DNT being released. Two loading schemes have been applied to trigger the pull-out process. One was to apply a constant velocity ($5\times10^{-5}$ Å/fs) to one end of the DNT, in which only the loading end was rigid and the rest part was flexible/deformable (Figure 1b). Such a loading scheme is similar to the steered MD simulation used to assess interfacial properties of polymers.[30] The other loading scheme applied a constant velocity ($1\times10^{-4}$ Å/fs) to the whole DNT while it was being treated as a rigid body (Figure 1c). Prior to the pull-out tests, the models were further relaxed under a constant energy (NVE) ensemble for 200 ps with the lateral edges of the polymer (thickness of 5 Å, Figure 1d) and the loading region of the DNT being fixed rigid. The pull-out simulation was then continued until the DNT was fully pulled-out. A low temperature of 100 K was adopted for all simulations to reduce the influence from the thermal fluctuations, and a time step of 0.5 fs was used throughout the pull-out simulation.

**Notes**

The authors declare no competing financial interests.

**Acknowledgement**

Supports from the ARC Discovery Project (DP130102120), the Australian Endeavour Research Fellowship, and the High Performance Computer resources provided by the Queensland University of Technology, Institute of High Performance Computing (Singapore) are gratefully acknowledged. In addition, we thank our colleague Mr. Edmund Pickering for proof reading.

**Supporting Information**

Supporting information is available for the energy change of the composite with DNT-4, further discussions of ISS and total vdW energy change trajectory of different composite models.

# Table of Contents

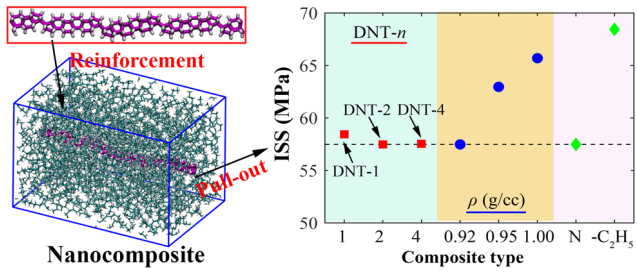